# Development of a multi-pixel photon sensor with single-photon sensitivity

Motohiro Suyama[1) 2)], Atsuhito Fukasawa[2)], Junji Haba[3)], Toru Iijima[4)]

Seigi Iwata[3)], Makoto Sakuda[3)], Takayuki Sumiyoshi[5)], Fumihiko Takasaki[3)],

Manobu Tanaka[3)], Toru Tsuboyama[3)], and Yoshikazu Yamada[3)]

1) Department of Particle and Nuclear Physics, The Graduate University for Advanced Studies, Oho 1-1, Tsukuba 305-0805, Japan

2) Electron Tube Center, Hamamatsu Photonics K.K., 314-5, Shimokanzo, Toyooka village, Iwata-gun 438-0193, Japan

3) Institute of Particle and Nuclear Studies, High Energy Accelerator Research Organization (KEK), Oho 1-1, Tsukuba 305-0805, Japan

4) Department of Physics, Nagoya University, Furou-cho, Chikusa-ku, Nagoya 464-8602,

5) Department of Physics, Tokyo Metropolitan University, Minami-Ohsawa 1-1, Hachiouji-shi, Tokyo,192-0364, Japan.

ABSTRACT

A multi-pixel photon sensor with single-photon sensitivity has been developed.  Based on a hybrid photo-detector (HPD) technology, it consists of a photocathode and a multi-pixel avalanche diode (MP-AD).  The developed HPD has a proximity-focused structure, where the photocathode and MP-AD face each other with a small gap of 2.5 mm.  The MP-AD, which has an effective area of 16mm×16mm, is composed of 8×8 pixel and has been specially designed for the HPD.  The maximum gain of the HPD is $5 \times 10^4$, sufficiently high to detect single photons with a timing resolution better than 100 ps.  Up to four photoelectrons can be clearly identified as distinct peaks in a pulse-height spectrum, thanks to the low noise characteristics of the HPD.  It is also demonstrated that the HPD can be operated with good performance in a magnetic field as high as 1.5 T.

Corresponding author: Motohiro Suyama

Postal address: Design 1 Group, Hamamatsu Photonics K. K., 314-5, Shimokanzo, Toyooka-village, Iwata-gun, Shizuoka-pref., 438-0193, Japan

Tel: 81-539-62-3151

Fax: 81-539-62-5005

e-mail: suyama@etd.hpk.co.jp

# 1. Introduction

In particle physics experiments, photon sensors have long played a key role in various particle detectors. One of the most interesting and attractive photon sensors in these days is a hybrid photo-detector (HPD). A HPD can be realized as combination of vacuum-tube technology and the rapidly evolving semiconductor technology. A photocathode and photodiode are mounted facing each other with a small gap inside a vacuum tube. The photoelectrons produced at the photocathode are accelerated across the gap by a strong electric field applied and impinge on the photodiode. The multiplication obtained through this acceleration can be as high as one thousand. Thanks to this high gain in the first multiplication stage, statistical fluctuations are so small that very clean photon counting is possible if a reasonable noise level can be maintained in the subsequent amplification stages. For a very fast application, however, it is still difficult to preserve good photon counting capability because of the relatively higher noise inherent to a high-speed amplifier.

This work focuses on the development of a new HPD equipped with a specially designed multi-pixel avalanche diode (MP-AD) in which fast secondary amplification will be achieved through an avalanche process with very low noise. Desirable features of the HPD, which we plan to develop, are as follows:

- Detection of single photons with a time response faster than one nanosecond
- A good signal-to-noise ratio to enable photon counting
- A multi-pixel structure with a pixel size of the order of 1 mm$^2$
- Operability in a magnetic field higher than 1 T.

In this paper, the construction of the developed HPD as well as its performance is described.

# 2. Structure of the HPD

A schematic drawing of the developed HPD is shown in Fig. 1. The primary components of the HPD are a photocathode, a MP-AD and a vacuum container.  The vacuum container consists of an input window, a cylindrical ceramic sidewall and a stem.  The ceramic sidewall has two openings, where the input window made of a fiber optic plate (FOP) seals one end, and the ceramic stem seals the other end to make a vacuum-tight structure.  The FOP is made of a bundle of optical fibers (6μm $\phi$) aligned to the tube axis so as to transfer an image focused on the front of the input window to the back without any loss of spatial information.  On the vacuum side of the FOP, antimony, sodium and potassium are evaporated to form a multi-alkali photocathode with an aperture of 25 mm in diameter.  The MP-AD is assembled on the stem facing the photocathode with a small distance of 2.5 mm to minimize the effect of any external magnetic field.  This proximity focusing structure is essential for vacuum tubes to be used in a high magnetic field.  Output pins come out for applying bias voltages or extracting signals in air back-side of the stem.

The MP-AD is specifically designed for this HPD.  Among the various types of avalanche photodiodes (APD) invented so far [1][2][3][4][5][6], the structure of the MP-AD shown in Fig. 2 was chosen taking the following points into consideration;

1) tolerance for intense electron bombardment in an HPD,
2) thin dead layer at the entrance,
3) applicability to a multi pixel structure,
4) ease of fabrication.

A Si substrate of 75 μm thickness was used so that full depletion may be achieved with a reasonable operational bias voltage, lower than 400 V.  Starting with a p type substrate, an 8×8 array of independent pn junctions, each of which forms a pixel, was fabricated on one surface while the other surface was treated for bombardment by electrons.  The pixel size of 2mm×2mm was chosen to have a capacitance around 6 pF, which corresponds to a timing response as fast as 1 ns for the HPD.  Since the pn junction is shielded by the substrate from direct hits by electrons, no damage is expected from electrons, assuring stable operation.  The highly doped p+ layer facing the electron flow is also useful to prevent charge accumulation due to electron irradiation.

## 3. Performance

### 3.1 Fundamental performance

The quantum efficiency (QE) of the photocathode is defined as the photon-to-electron conversion efficiency, which is a function of wavelength. As shown in Fig. 3, the QE of the present HPD is measured to be 20% in the UV and blue region while it is 10% or less in the visible region, a typical response for a multi-alkali photocathode. The cutoff wavelength of 360 nm is determined by the transmission characteristic of the FOP.

The electron-bombarded gain (EB gain) defined as the ratio of the currents from the MP-AD and the photocathode, was measured as a function of the photocathode voltages. As shown in Fig. 4, the EB gain starts rising around a threshold voltage ($V_{th}$) of -4.0 kV and reaches 1270 for a photocathode voltage of -8 kV. The EB gain ($m_{EB}$) can be approximated by the following equation:

$$m_{EB} = \frac{-(V - V_{th})}{E}, \qquad (1)$$

where $V$ is the photocathode voltage, $V_{th}$ is the threshold voltage, and $E$ is the average energy needed to produce one electron-hole pair. $E$ can be estimated from the slope of the gain to be 3.2 eV. This is close to the theoretical value of 3.6 eV for silicon [7]. The threshold energy is thought to be determined by the thickness of the dead layer at the entrance surface. The actual thickness was evaluated by modeling the behavior of EB gain with the PENELOPE [8], simulation code for electron/photon transport. The best fit, shown in Fig. 4, was obtained with a dead layer of 210 nm, which approximately corresponds to the thickness of the p+ layer. Disagreement seen at the threshold can be attributed to such a simple assumption that any electrons generated in the dead layer can never reach the avalanche region. Although the p+ layer is absolutely necessary to prevent thermal charges from flowing out as an excess dark current, there is still room to optimize its thickness. The maximum voltage on the photocathode was limited below 9kV in the present test to avoid discharges between the photocathode and AD, which sometimes damages the MP-AD badly. Some of the measurements in the early period like the linearity measurement were done with an even lower voltage for a safer operation.

The V-I characteristic, the leakage current of the MP-AD as a function of the reverse bias voltage (AD voltage), was measured in a dark environment. The results are shown in Fig. 5, where the current was summed from all pixels. It is seen that full depletion is realized at the kink around 100V.

The avalanche gains were measured under continuous illumination by photons or electrons and plotted in Fig. 6, where the currents are normalized to those measured at the AD voltage of 1V assuming the avalanche gain to be unity there. The solid line shows the avalanche gain for illumination by photons. The dashed and the dotted lines illustrate those for electrons with photocathode voltages of -6 kV and -9 kV, respectively. It was found that the avalanche gain showed a slight decrease when the photocathode voltage was increased. One possible explanation is a local space-charge effect at the avalanche multiplication region. A primary electron from the photocathode generates approximately one thousand electrons inside the AD, localized in a volume of several cubic microns. The localized charge compensates the electric field at the avalanche region and, therefore, reduces the avalanche gain significantly. Since the space-charge effect is quite localized, it does not seriously affect the linearity characteristic for counting of primary photoelectrons. Good linearity at least up to 230 photoelectrons per pixel was confirmed, as shown in Fig. 7. This was measured by the setup described in section 3.2, where the number of incident photons was varied using neutral density filters. The total gain is $5.0 \times 10^4$ at the photocathode voltage of –8 kV with 340 V on the MP-AD, just below the threshold of an excessive leakage current. No significant deviations from linear response were observed in the reachable range of the present measurement.

### 3.2  Pulse-height spectra

The pulse-height spectra for single photons were measured for various photocathode voltages using a light emitting diode (LED) of 470 nm wavelength as a pulsed light source.  The HPD output signal was amplified by a charge sensitive amplifier (ORTEC, 142A [9]), further amplified and shaped by a linear amplifier (CANBERRA, 3100-02 [10]), and then analyzed by a multi-channel analyzer (Laboratory equipment corp., MCA/AT [11]).  Figure 8 shows the measured pulse-height spectra for photocathode voltages of -9 kV (solid), -8 kV (dashed) and -7 kV (dotted), respectively, with the AD voltage fixed at 250 V.  The total gains for these conditions were $2.7 \times 10^4$, $2.1 \times 10^4$ or $1.6 \times 10^4$, respectively.  The resolution ($R$), which is defined as the ratio of the full width at half maximum (FWHM) to the peak channel of the spectrum, improves with increasing photocathode voltage, as shown in Fig.8.

The improvement of resolution with photocathode voltage can be explained in terms of the energy loss of electrons in the dead layer at the entrance surface of the MP-AD.  When electrons from the photocathode lose a significant part of their energy in the dead layer, the energy detected in the sensitive layer would experience large event-by-event fluctuations.  Accordingly, the EB gain fluctuates.  The simulation code, PENELOPE [8], was used to simulate the energy loss and fluctuations in the dead layer.  Based on the result from the EB gain, the thickness of the dead layer was assumed to be 210 nm.  The computed fractional energy loss and its fluctuation (in FWHM) to the deposited energy in the sensitive layer are shown in Fig. 9.  The energy losses were evaluated to be 2.9 keV (41% of the incident energy) at the photocathode voltage of -7 kV, 2.5 keV (31%) at -8 kV and 2.2 keV (25%) at -9 kV, while the fluctuations (in FWHM) of the energy loss were calculated to be 44% at -7 kV, 25% at -8 kV and 18% at -9 kV.  The smaller fluctuation in EB gain for higher energy electrons results in better pulse-height resolution.

The pulse-height spectra for single photons were also measured as a function of the AD voltages with the photocathode voltage fixed at -8 kV. The solid, dashed and dotted lines in Fig. 10 show spectra for AD voltages of 340V, 312V, and 250V, respectively. The increase in the peak channel with the AD voltage can be simply explained by the increase of the avalanche gain. The resolution is improved with increasing AD voltage, as shown in Fig. 11. This may be explained by the behavior of the avalanche gain as a function of EB gain, as shown in Fig. 6. Events with higher (lower) EB gain due to fluctuations tend to have more (less) saturation in the avalanche multiplication process, and thus produce a relatively lower (higher) gain there. The compensation could be more significant for higher AD voltage. Accordingly, the overall gain may fluctuate less and the pulse-height resolution be better for higher AD voltage.

Using the same setup, the photon intensity was increased and the pulse-height spectra for multi-photon signals were studied with the maximum applicable voltages to the AD (340V) and photocathode (-9kV). As shown in Fig. 12, the peaks corresponding to 1 to 4 photoelectrons are clearly identified. This demonstrates the low noise characteristics of the developed HPD.

### 3.3 Time response

The time response for single photons was measured with an oscilloscope in the averaging mode. The LED light intensity was set to be single-photon level. The HPD was operated at a gain of $4.2 \times 10^4$ with moderate voltages on the photocathode and the AD of –8kV and 325V, respectively. The time response of the HPD was found insensitive to the variation of the voltage on the photocathode or on the AD in the range of 250V to 340V. A high-speed amplifier (Miteq, AU-1494-300, 300 MHz, Gain: 60 dB [12]) was used to amplify the output signal from the HPD. The captured trace as displayed in Fig. 13 yielded rise and decay times of 1.1 ns and 1.3 ns, respectively. The noise level was measured to be 13 mV in rms with the oscilloscope in the real time mode, which should be compared with a single photon signal amplitude of 240mV. The signal to noise ratio (S/N) is thus calculated to be as high as 18 for a single photon detection. This low noise characteristic and the high speed response have never been achieved by any standard HPD without avalanche gain.

The timing resolution for a single photon was then measured with a pulsed laser diode (LD), with a pulse width of 67 ps in FWHM and a wavelength of 405 nm. After a two-stage amplification (Miteq: AU-1494-300 [12] and HP8447F: 1.5 GHz [13]), the output signal from the HPD was fed to a constant fraction discriminator (CFD, ORTEC 583 [9]) to generate the start signal for a time-to-amplitude converter (TAC, ORTEC 547 [9]). The trigger signal from the LD driver was used as the stop signal of the TAC. The TAC output was recorded by the MCA. The distribution of measured time intervals at the gain of $5.7 \times 10^4$ is shown as open circles in Fig. 14. By fitting with a Gaussian function (solid curve), the timing resolution for a single photon is estimated to be 80 ps including the timing jitter of the measurement system (26 ps) and light pulse (29 ps). The resolutions at lower gain are also measured to be 97 ps ($4.7 \times 10^4$) and 104 ps ($4.2 \times 10^4$), confirming the dependence of the timing resolution upon the gain of the HPD, as discussed in [14]. The resolution is compatible with that of the transit time spread (TTS) of commercially available fast PMTs.

### 3.4  Performance as a multi-pixel photon sensor

As a multi-pixel device, uniformity, cross-talk and response at the pixel boundary were investigated in detail. The uniformity over the HPD was measured by scanning the light spot of 0.94 mm in diameter focused on the center of every pixel. The wavelength of the light was chosen to be 500 nm. The HPD was operated at a gain of $4.2 \times 10^4$ with moderate voltages on the photocathode and the AD of –8kV and 325V, respectively. This operational condition was used for the measurements hereafter. The relative amplitude of the measured output from each pixel is plotted in Fig. 15. The standard deviation of the relative gains among the 64 pixels was found to be 7.9% with respect to the average one, and the smallest output was 75% of the largest. This uniformity is better than that of 30% as reported for a multi-anode PMT commercially available in the market [15].

To identify the source of the non-uniformity, further investigations on the photocathode sensitivity and EB gain were made. To measure the photocathode sensitivity, the cathode and anode of the AD were short-circuited, and the HPD was operated as a kind of phototube. With -1 kV on the photocathode, the output current was measured. To examine the uniformity of EB gain, the HPD was operated with a unit avalanche gain in the AD and its output was normalized by the photocathode sensitivity measured above. As a result, the non-uniformity in the photocathode sensitivity and EB gain are measured to be as small as 1.5% and 1.3%, respectively, which can not account for the observed overall non-uniformity. The uniformity in the avalanche gain was evaluated by dividing the HPD output under normal conditions by the output with the unity gain in the avalanche diode. As seen in Fig. 16, the overall gain is fully correlated with that of the avalanche diode. It can be concluded that the non-uniformity of the HPD is dominated by that of avalanche multiplication in the MP-AD. This can be attributed to a slight gradient in the impurity density in the p layer facing the n+ layer. The prior test with the prototype AD shows that a 2% decrease in impurity density reduced the avalanche by a factor of 2 because of the lower electric field in the avalanche region.

The cross-talk behavior between pixels was studied under continuous illumination focused onto the center of a single pixel with a spot size of 0.94 mm. The output currents observed on the other pixels were measured as shown in Fig.17. It is seen that the cross-talk into the adjacent pixel is 1.3% while it is much smaller for the others. Some of the cross-talk is considered to come from the light scattering in the FOP. The electrons backscattered at the MP-AD also contribute, as discussed later.

The cross-talk between pixels under pulsed light was also investigated. A spot of pulsed light was focused onto the center of pixels, and the output charge appearing on other pixels was measured by a charge-sensitive amplifier (ORTEC 142A [9]). As shown in Fig. 18, the cross-talk characteristic was almost the same as that observed under continuous illumination, except for two neighboring pixels (pixels (5, 6) and (5, 7)) whose signal due to cross-talk was found to be approximately 6%. The latter can be attributed to capacitive coupling among the signal traces on the AD, where they run side-by-side with a separation of 0.2 mm. This also explains why the cross-talk to pixel (5,3) or (5,4) was small, whose signal trace runs toward the opposite direction. The cross-talk can be reduced by modifying the signal traces to have larger separation.

The gain characteristic at the edge of the pixel was examined by scanning across the boundary of two adjacent pixels with a light spot of 30 μm diameter. As shown in Fig. 19, there is a low gain zone 120 μm wide where the gain falls less than 50% of that in the center of the pixel. The dashed line drawn in the figure indicates the expectation assuming both a low gain zone of 120 μm wide with 25% gain and finite size of the irradiated light spot.

**3.5  Operation in a strong magnetic field**

Several characteristics of the HPD were also measured in a magnetic field. The input FOP window was masked to leave an opening of 1.0mm×1.2mm to restrict the illumination from the LED (470 nm) onto a single pixel. The HPD was placed inside the electromagnet, aligning its axis at $0^o$ and $30^o$ with respect to the magnetic field direction. The output current from the illuminated pixel and an adjacent one were monitored while changing the field strength from 0 to 1.5 T. The experimental configuration is illustrated in Fig. 20.

The output signal observed in the illuminated pixel and the adjacent one are plotted in Fig. 21 as a function of the field strength. It is seen that the output in the illuminated pixel increases by 10 % from 0.1T to 0.6 T, while that in the adjacent pixel decreases from 1.4 % to 0.8%. This can be understood as re-entry of backscattered electrons onto the illuminated pixel under a strong magnetic field. These electrons would otherwise be distributed to adjacent pixels and produce the substantial cross-talk signal reported in the previous section. The amount of distributed charge over the 24 surrounding pixels can account for the 10% loss in the illuminated pixel at zero field.

Where the magnetic field was inclined by 30 $^o$ to the tube axis, the output current from the illuminated pixel rapidly decreased with the field strength, while a corresponding increase was observed in the adjacent pixel. This can be explained by a Lorenz shift of the trajectory of photoelectrons. A calculation of the trajectory in the magnetic field shows that the position of the electrons on the MP-AD is shifted by 1.43 mm for a magnetic field higher than 0.5 T so that the electron hits the adjacent pixel

## 4. Practical test

In order to demonstrate the capability of the HPD in a particle detector, it was integrated with a bundle of scintillating fibers (Kuraray SCSF-78M, 0.7mm$\phi$), which was directly connected to the FOP window of the HPD. The peak wavelength in the emission spectrum of the fibers was 430 nm. The length of the fiber bundle was 200 mm and the cross section was formed to 16mm×16mm to match the effective area of the HPD. A mirror was put on the open end of the scintillating fibers to reflect photons towards the HPD. The applied voltages were -8 kV on the photocathode and 315 V on the AD for the gain of $3.8 \times 10^4$. All 64 pixels of the HPD were connected to readout circuits, which consisted of a 64 input front-end chip (IDE, VA32_HDR2 [16]) with a shaping time of 1μs, control circuitry and an analog-to-digital converter (ADC) board (National instruments, PCI-6111E [17]) mounted on a personal computer. The lower AD voltage was chosen to reduce the leakage current flowing into the multi-channel front-end chip connected directly to the AD outputs.

Cosmic-ray tracks were recorded with this system, as shown in Fig. 22, where the shading in the squares correspond to the number of photoelectrons detected in the pixels. The average number of photoelectrons per pixel is estimated to be 17. This may be compared with the expectation of 7.4 photoelectrons deduced from the past measurement [18], where the significantly longer fibers (3700mm) were used. It is also seen that the noise level is well below that of a single photoelectron. Thus, it is demonstrated that the HPD developed in the present study is realistically applicable for photon-counting in particle detectors.

## 5. Summary and future prospect

We have successfully developed a new type of HPD which is composed of a photocathode and a multi-pixel avalanche photodiode (MP-AD) specifically designed for this application. The overall gain as high as $5\times10^4$ was achieved with a timing resolution of 80 ps for a single photon. This high gain and precise timing resolution have been achieved for the first time in an HPD thanks to a suitably designed MP-AD. The pixel-to-pixel non-uniformity in the sensitivity was found to be as small as 8 %, which is substantially better than that of any multi-anode PMTs that are commercially available. The new HPD also shows reasonable performance in a magnetic field as high as 1.5 T. The HPD was applied to particle tracking in conjunction with a scintillating fiber array, and cosmic-ray tracks were observed with good signal to noise ratio.

There are still some aspects, which can be further improved. One is the arrangement of pixels. The technology for an MP-AD developed in this study can be adapted to various applications with simple geometric modifications. A pixel arrangement with 32×32 pixel in the same area of 16mm×16 mm exists, although the low gain zone between pixels is not negligible (0.1mm wide) in the present prototype. Reduction of the low gain zone should be investigated further by optimizing the inter pixel structure of the avalanche region. The effective area of the current HPD is limited by the size of the vacuum container, which is optimized for a photocathode diameter of 25 mm. The MP-AD was designed to match the size of the photocathode.  It is technically possible to apply a larger AD for a larger vacuum container.  A number of output pins may limit a number of pixels less than 100 for realizing a stem.  In the case of more than 100 pixels, it is preferable to encapsulate a chip for multiplexing output signals from an MP-AD.

A proximity focused HPD that is less sensitive to magnetic fields may be constructed by reducing the distance between the photocathode and the MP-AD.  In this case, it is necessary to avoid voltage breakdown in the higher electric field due to the reduced distance between the photocathode and the MP-AD.  One possible solution is a thinner dead layer of the entrance surface of the MP-AD so as to generate the same EB gain with a lower voltage to the photocathode.

The other aspect to be improved is the drastic enlargement of the effective area by using an electrostatically focusing lens.  In this case, electrons from a large-area photocathode are demagnified and focused onto an MP-AD by an electron lens.

**Acknowledgement**

We wish to thank Prof. F. Salvat of Universitat de Barcelona for instruction in the simulation code PENELOPE, and useful discussions about the results.  We would like to express our thanks to Messrs. K. Yamamoto, Y. Ishikawa and M. Muramatsu of the solid state division of Hamamatsu Photonics K.K. (HPK) for technical discussions about avalanche diodes.  We greatly appreciate the efforts of Messrs. A. Kageyama and K. Inoue of the electron tube center (ETC) of HPK for assembling the avalanche diode, and Mr. Y. Kawai of the ETC of HPK for discussions about its evaluation.  We are also grateful to Prof. Kay Kinoshita of Cincinati University who kindly read and improved our manuscript.  This work was partially supported by Grant-in-Aid for Scientific Research (B) (2) 13554008 of the Japanese Ministry of Education, Culture, Sports, Science and Technology.

**[Figure Captions]**

Fig. 1: Schematic illustration of the developed HPD.

Fig. 2: Structure of the multi-pixel AD (MP-AD) in the HPD.

Fig. 3: Quantum efficiency of the HPD as a function of the wavelength.   The cutoff wavelength is determined by the transmission characteristic of the FOP window.

Fig. 4: Measured and simulated electron-bombarded gain as a function of the photocathode voltage. The extrapolated line is expressed by the equation 1 given in the main text.

Fig. 5: Leakage current of the AD as a function of the AD voltage, where the currents from all pixels are summed.

Fig. 6: Avalanche gain as a function of the AD voltage for various photocathode voltages.   The solid line shows the avalanche gain for light input.   The dashed and the dotted lines illustrate those for electrons with the photocathode voltage of -6 kV and -9 kV, respectively.

Fig. 7: Output pulse height in the unit of photoelectrons as a function of input light intensity.   The solid line is drawn for a perfect linear response constrained at the intensity of 0.05.

Fig. 8: Single-photon spectra for various photocathode voltages, with the AD voltage fixed at 250 V. The solid, the dashed and the dotted lines show spectra for photocathode voltages of -9 kV, -8 kV and -7 kV, respectively.

Fig. 9: Computed fractional energy loss of electrons in the surface dead layer and its fluctuation (in FWHM).   The thickness of 210 nm was assumed for the dead layer, assuming to the result obtained in Fig. 4.

Fig. 10: Single-photon spectra for various AD voltages, with the photocathode voltage fixed at -8 kV. The solid, the dashed and the dotted lines show spectra for AD voltages of 340V, 312V and 250V, respectively.

Fig. 11: Pulse-height resolution for single photons, as a function of the AD voltage.

Fig. 12: Pulse-height spectra for multi-photons.

Fig. 13: An averaged output signal trace from the HPD for single photons recorded on an oscilloscope.

Fig. 14: Timing resolution for single photons (open circles). The solid line shows the fit by a Gaussian distribution with σ of 80 ps.

Fig. 15: Gain uniformity measured by scanning a light spot on the center of each pixel.

Fig. 16: Correlation between the avalanche gain and the overall gain for each pixel.

Fig. 17: Cross-talk measured with a continuous light.

Fig. 18: Cross-talk with a pulsed light.

Fig. 19: Response at the boundary of the pixels measured by scanning a light spot. Also drawn with a dashed curve is the expected response assuming that a square low-gain zone (25% of normal part) of 0.12 mm wide is smeared by the finite light spot size of 30 μm.

Fig. 20: Schematic illustration of the test in a magnetic field. The coordinate as well as angle (θ) of the magnetic field to the tube axis are given.

Fig. 21: Output currents in the illuminated and the adjacent pixels as a function of the strength of a magnetic field. The currents are normalized to that of the illuminated pixel under 0 T. The closed circles and triangles represent for illuminated and its adjacent ones, respectively, with the magnetic field aligned to the HPD axis, while the open circles and triangles for the case of 30 degrees inclination. The closed triangles are plotted in the scale given on the right axis, while other symbols on the left one.

Fig. 22: Passages of cosmic ray in the bundle of scintillation fibers viewed by the HPD.

**[REFERENCES]**

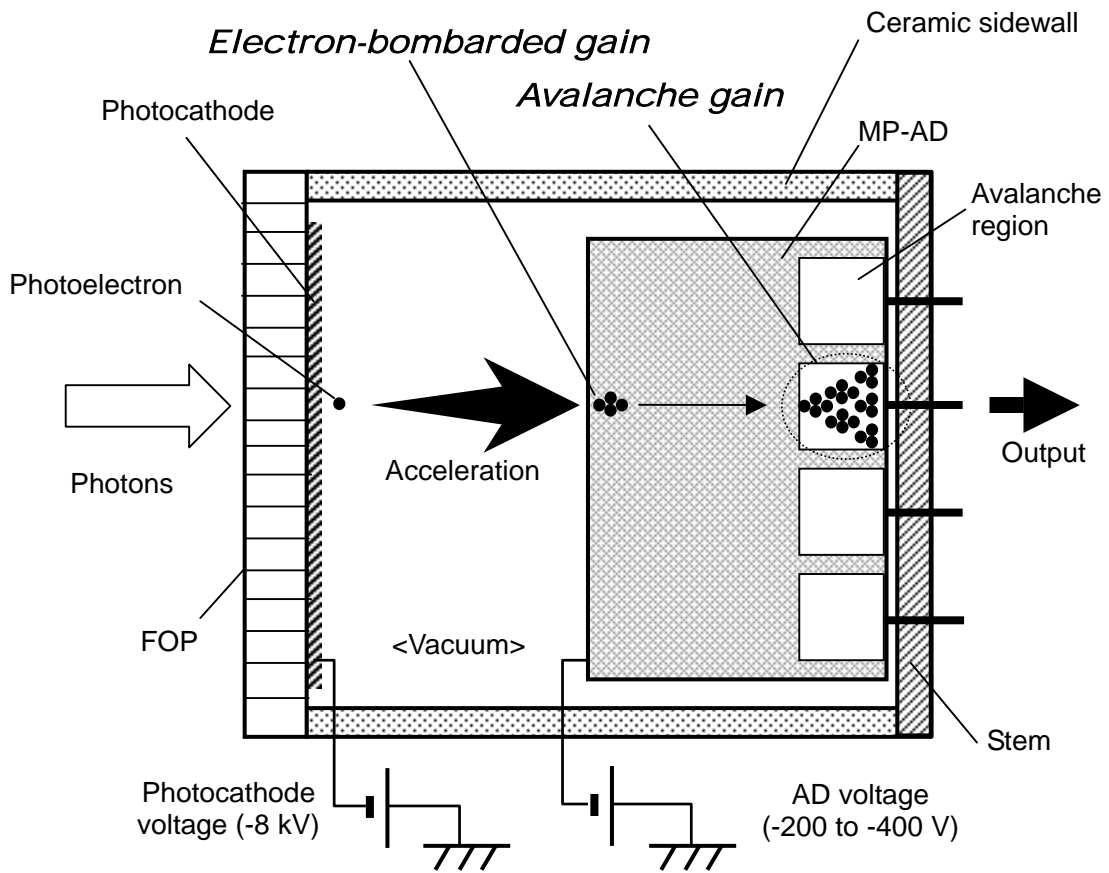

**Fig. 1**

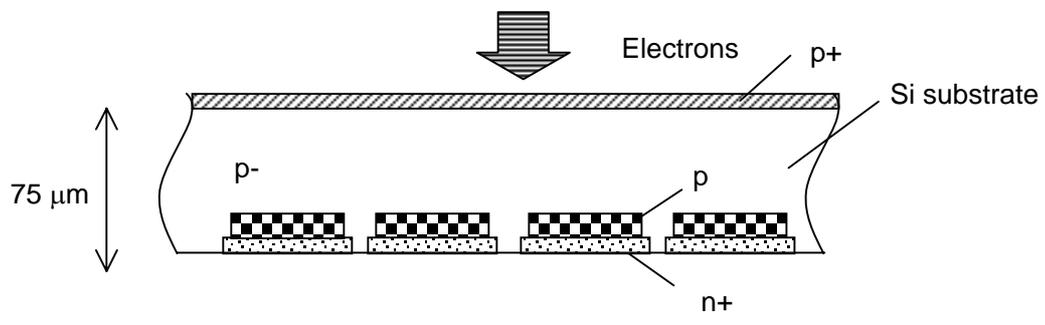

**Fig. 2**

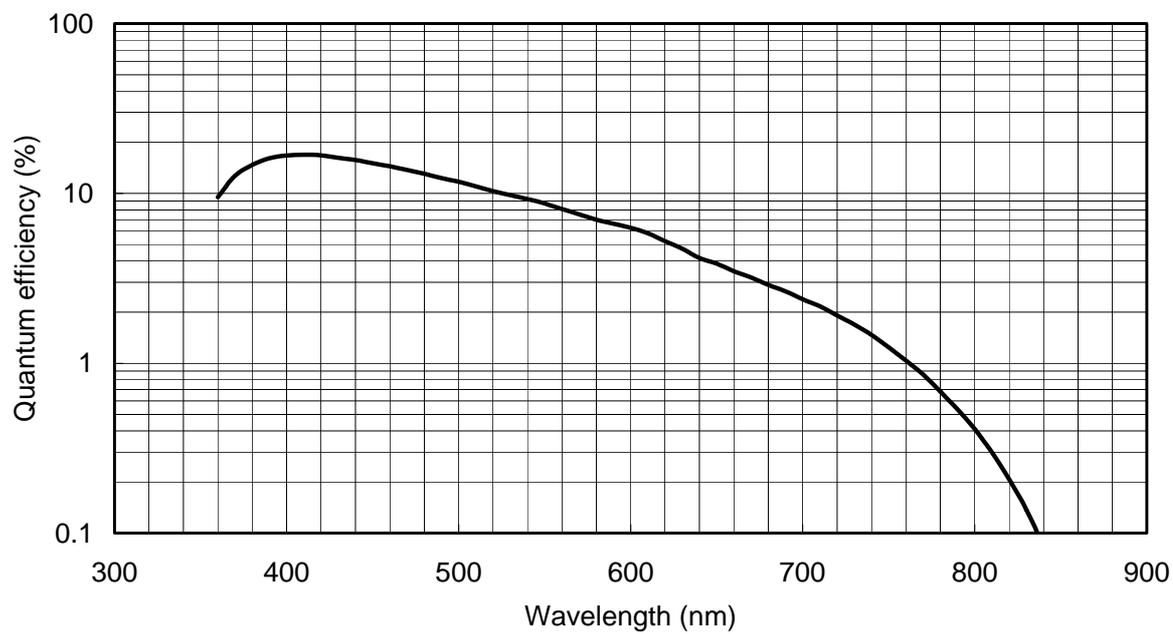

**Fig. 3**

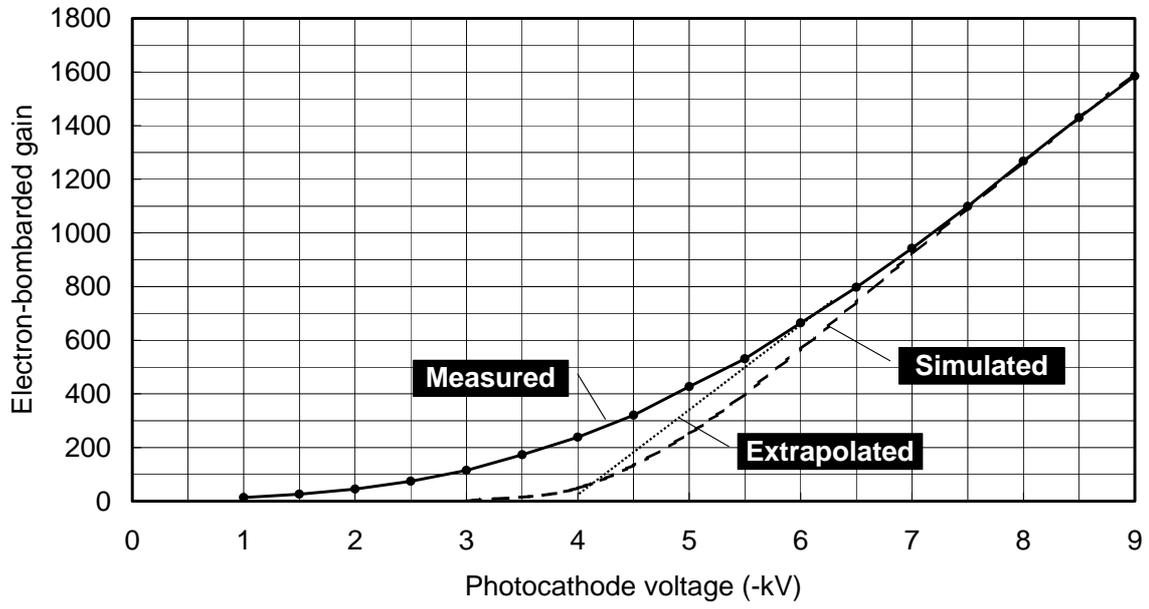

**Fig. 4**

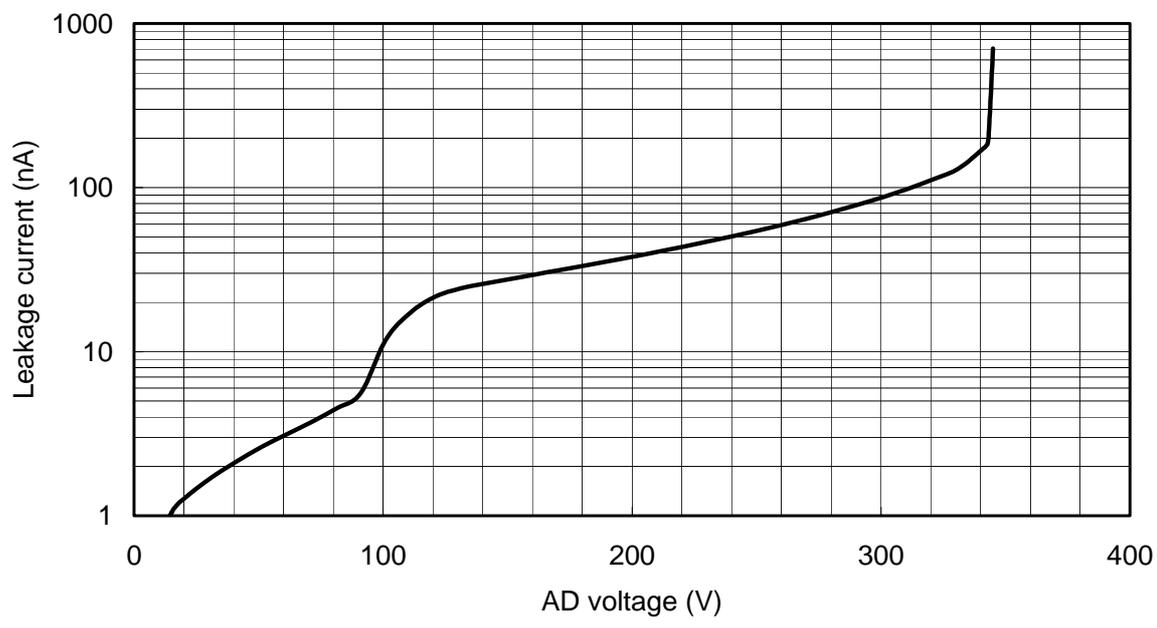

**Fig. 5**

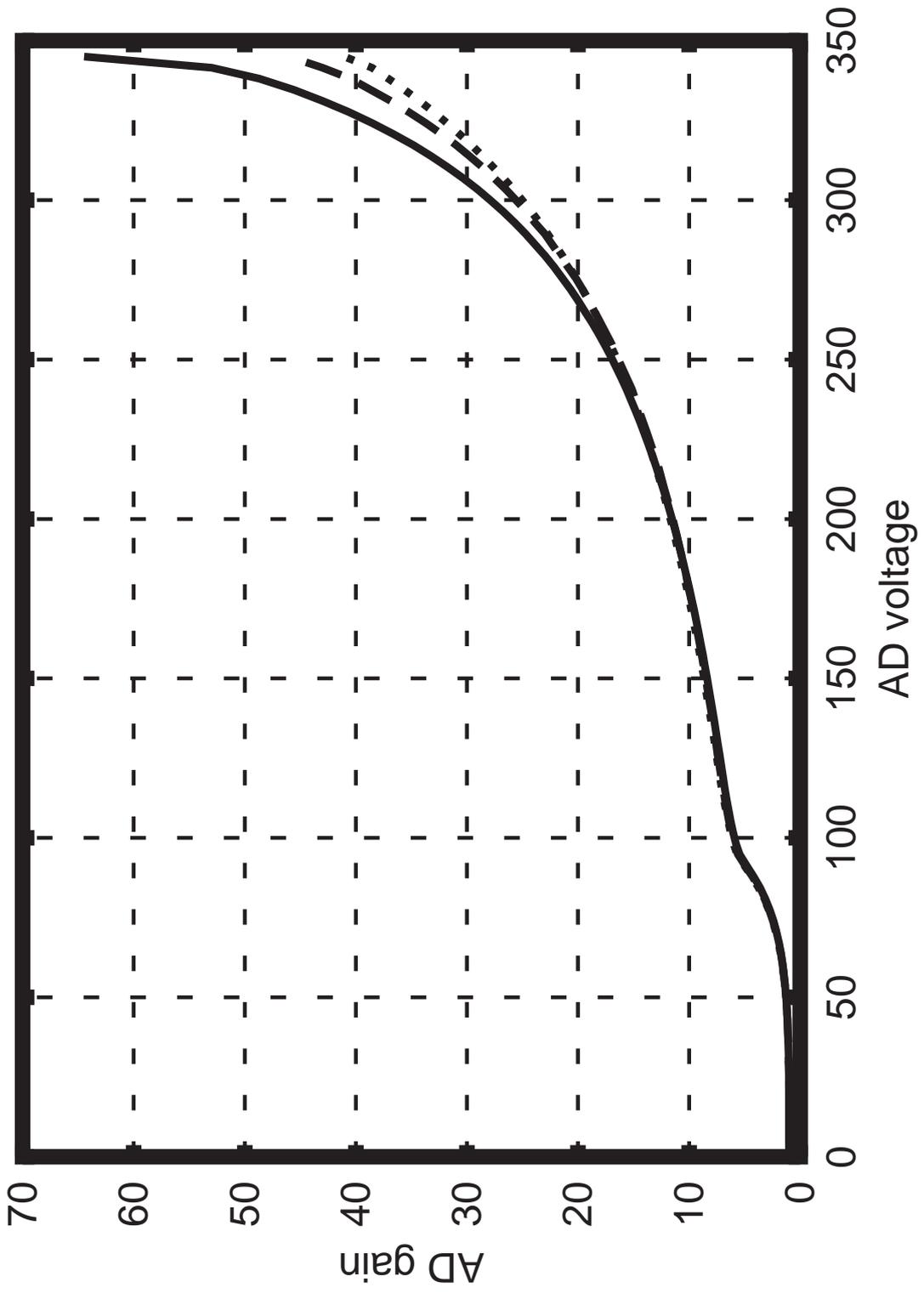

**Fig. 6**

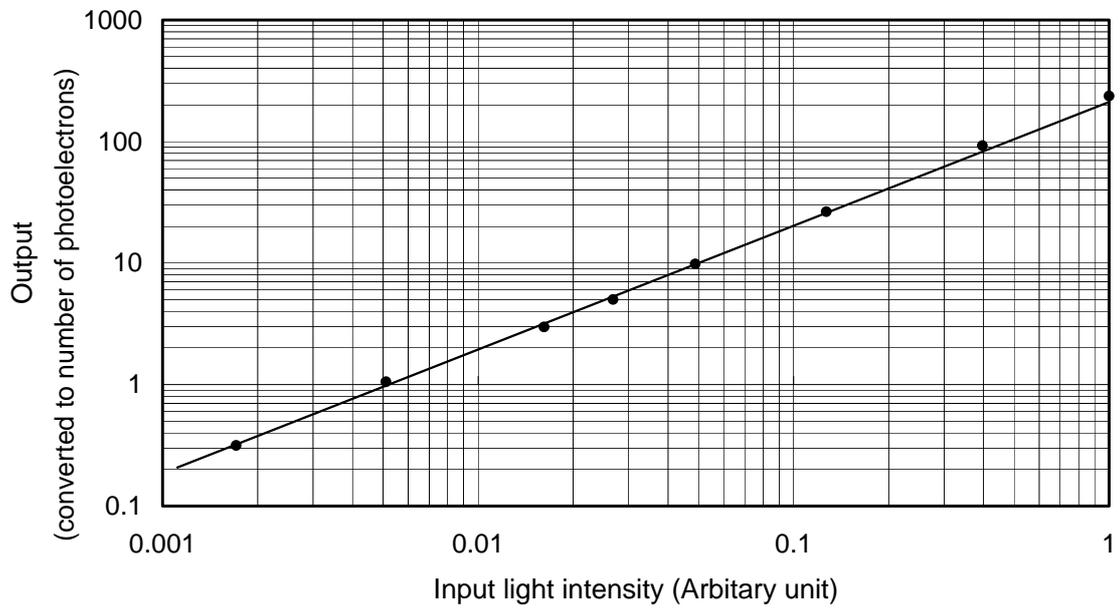

**Fig. 7**

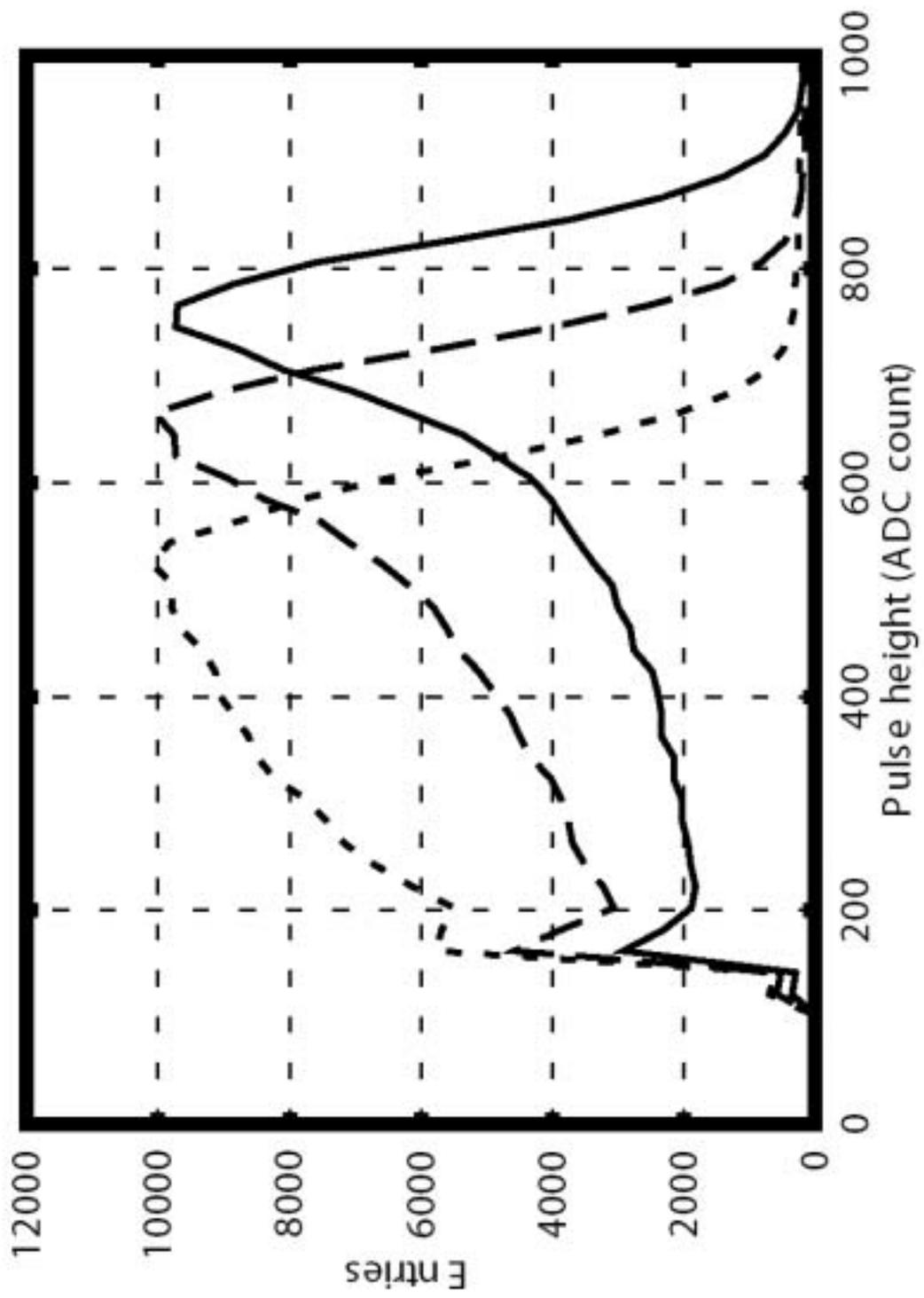

**Fig. 8**

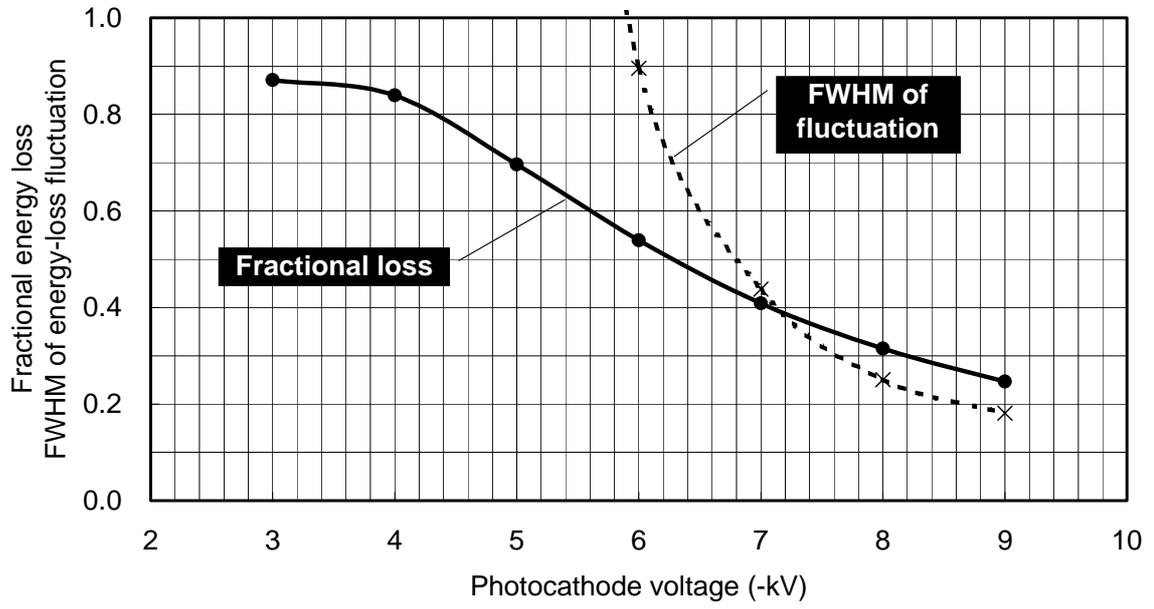

**Fig. 9**

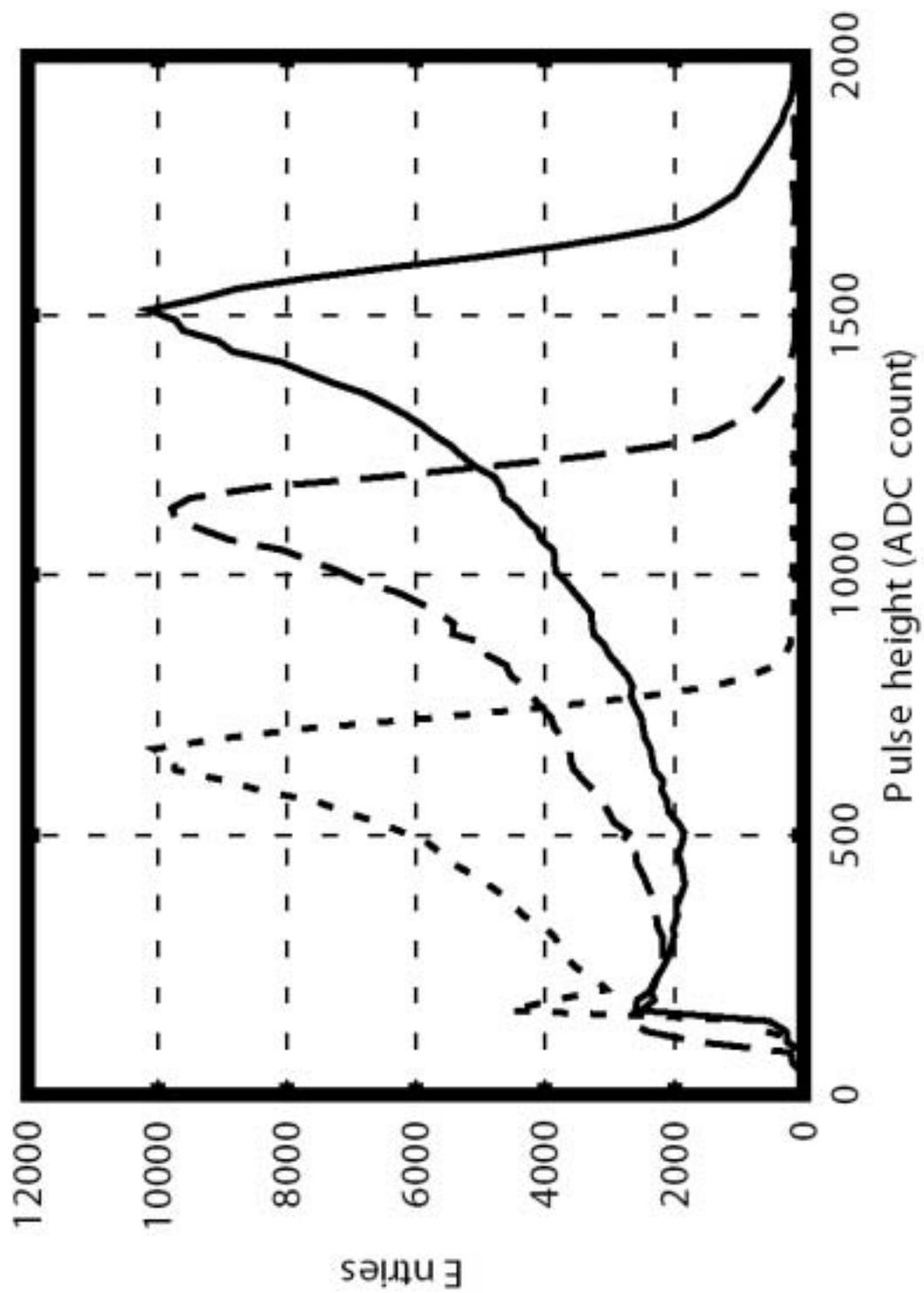

**Fig. 10**

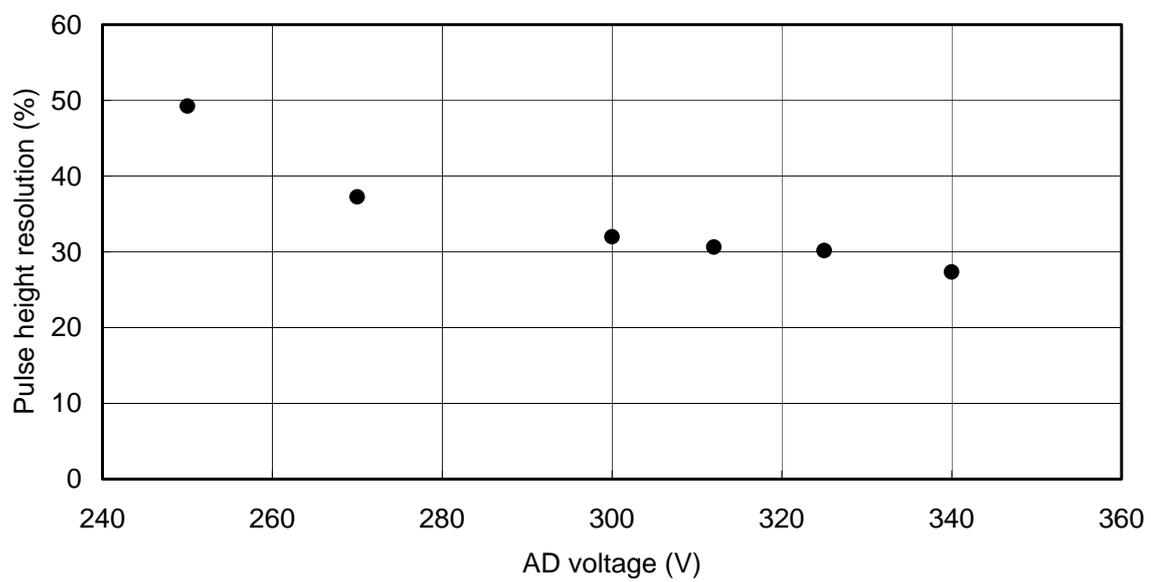

**Fig. 11**

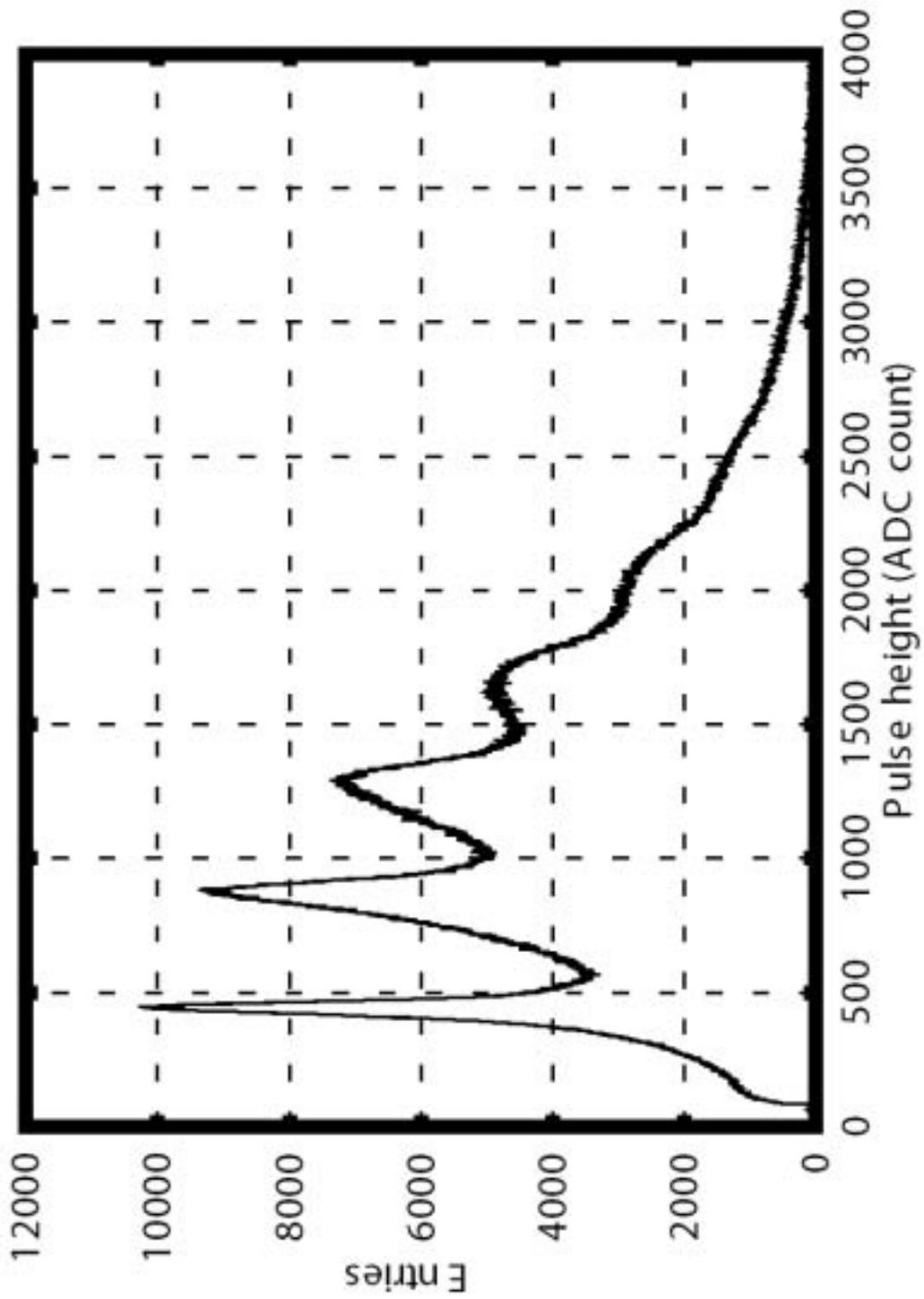

**Fig. 12**

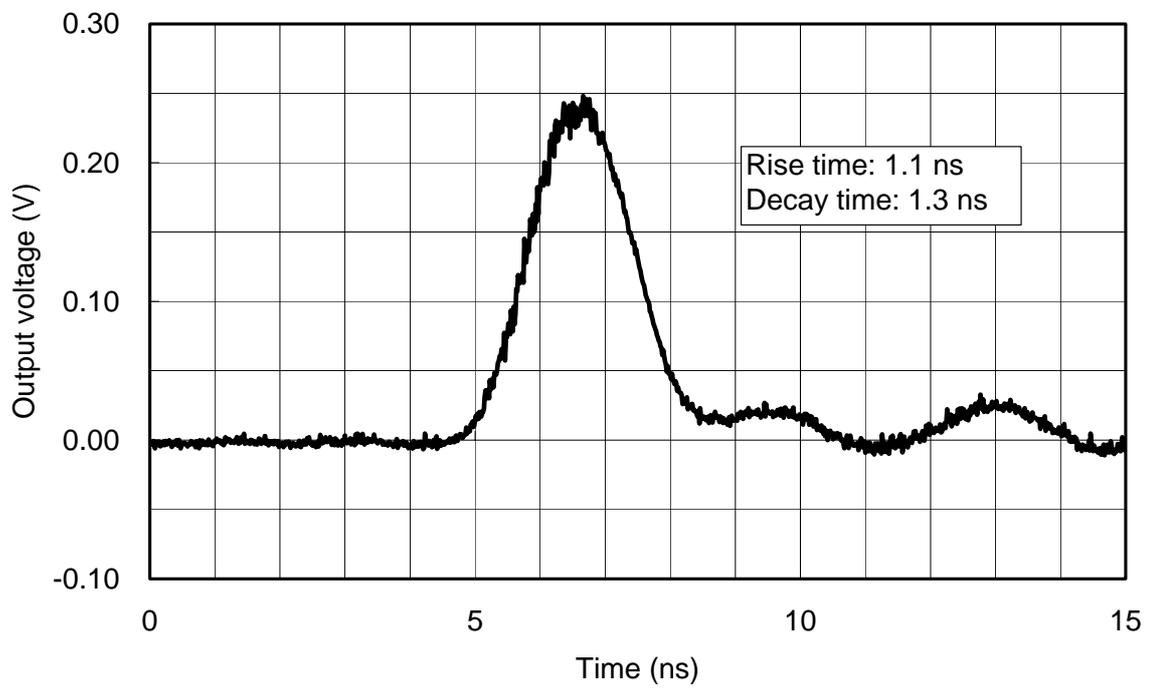

**Fig. 13**

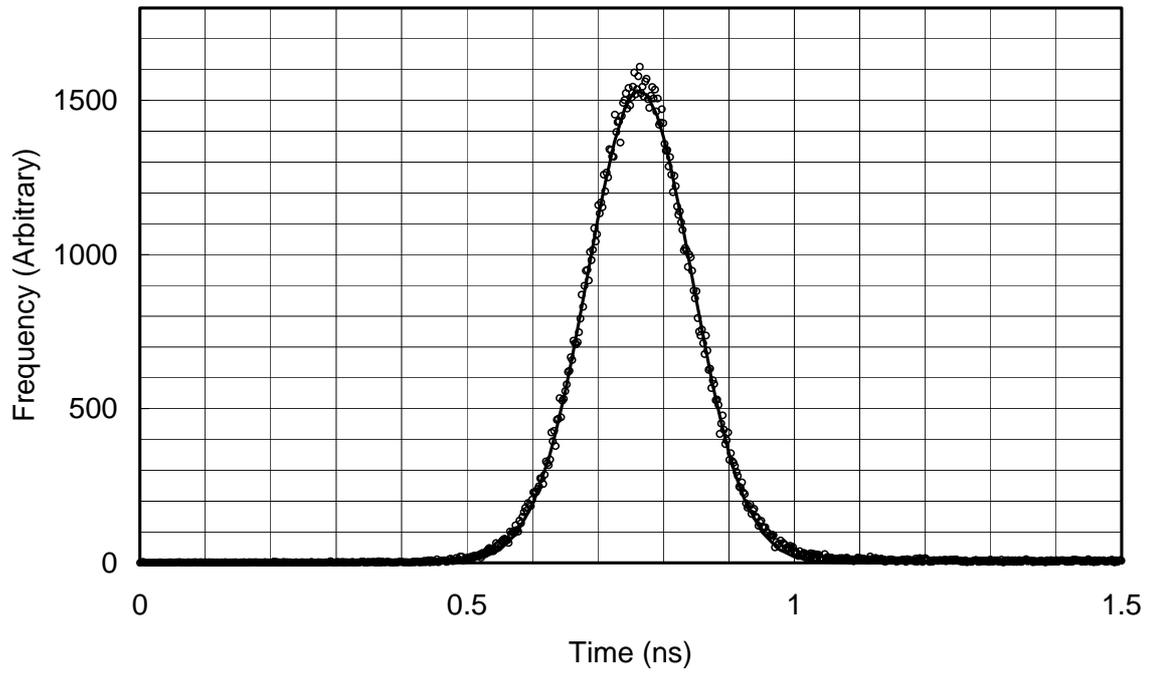

**Fig. 14**

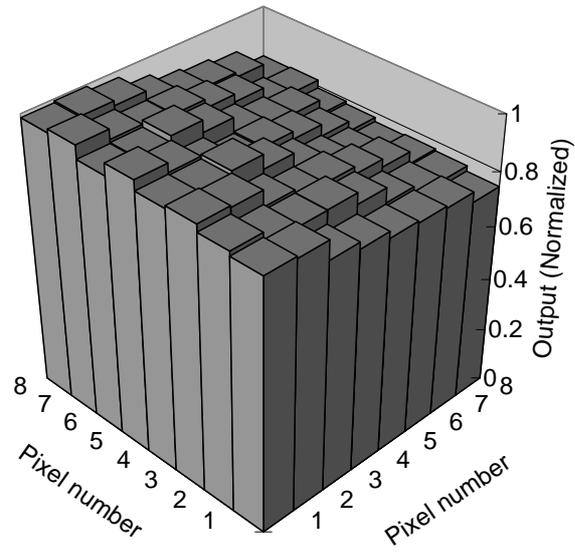

**Fig. 15**

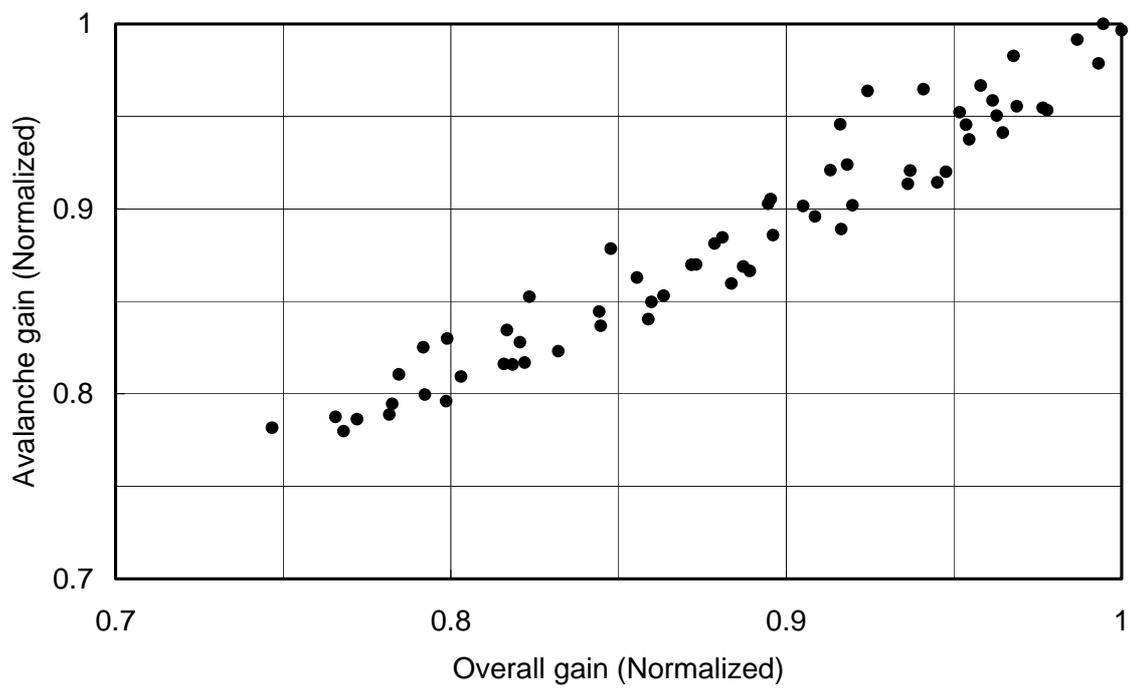

**Fig. 16**

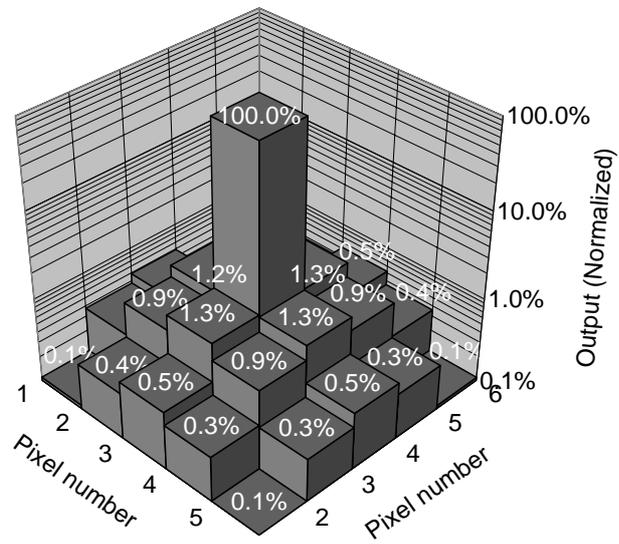

**Fig. 17**

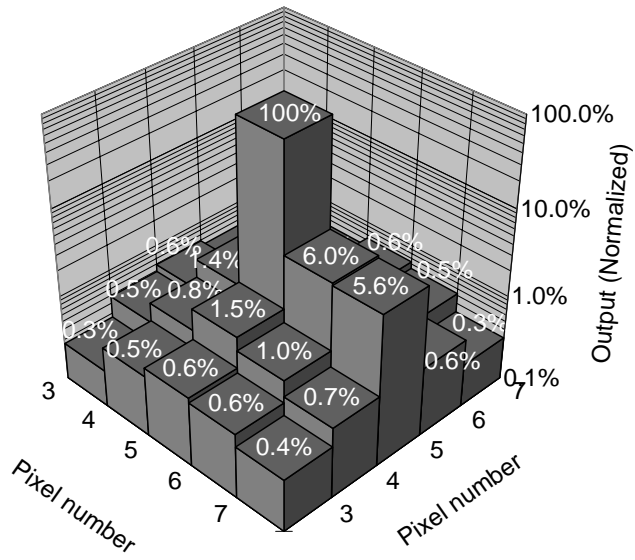

**Fig. 18**

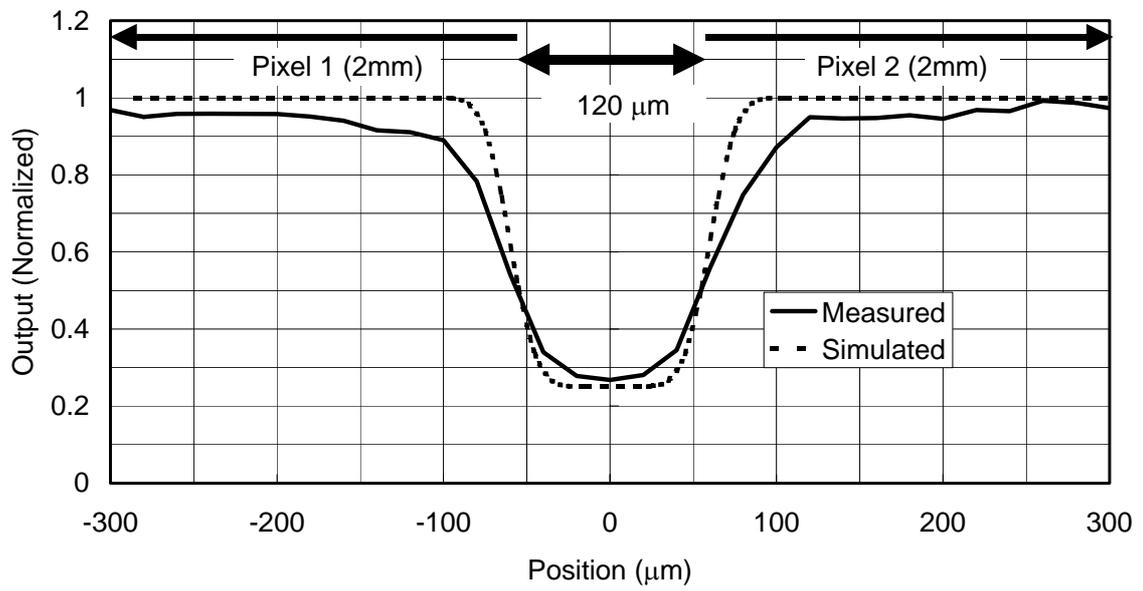

**Fig. 19**

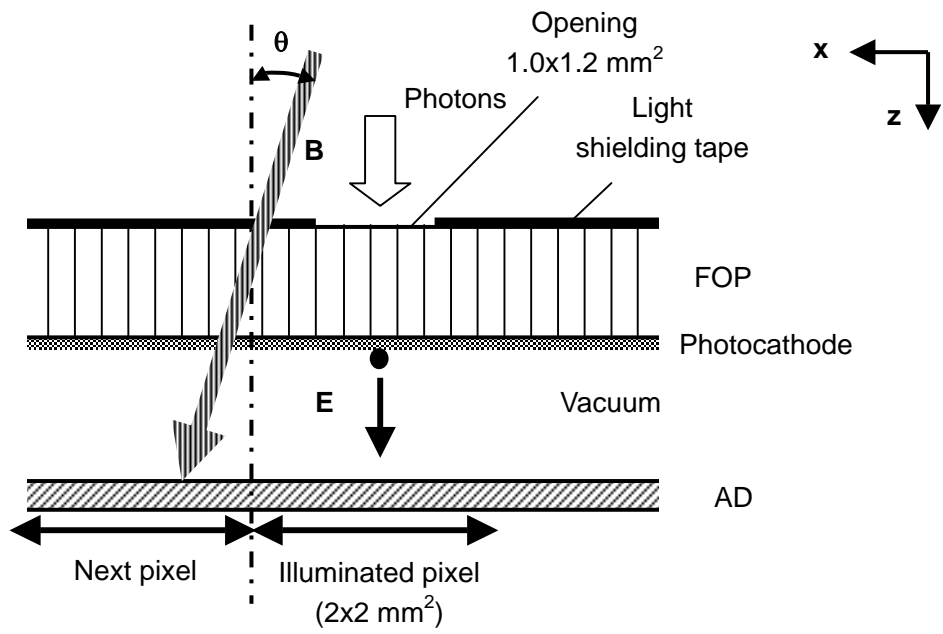

**Fig. 20**

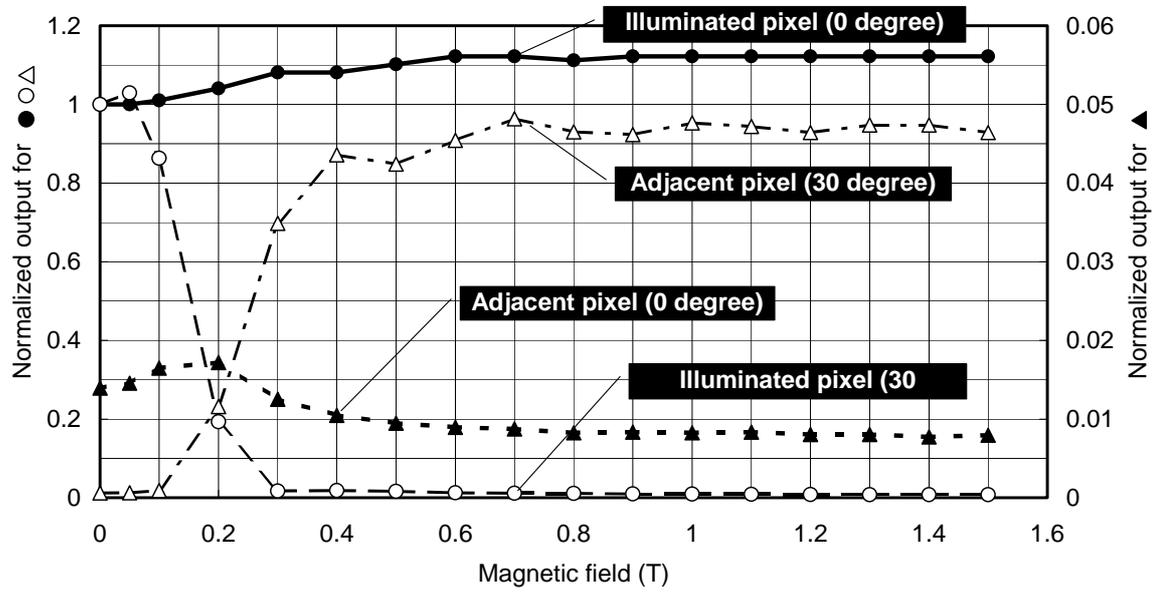

**Fig. 21**

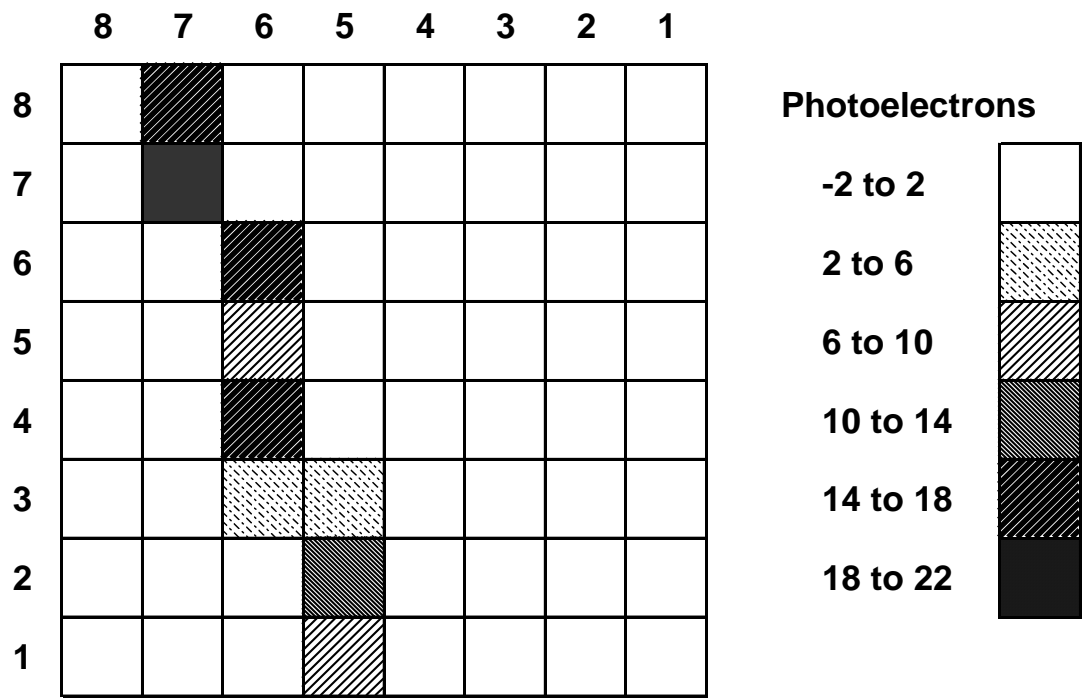

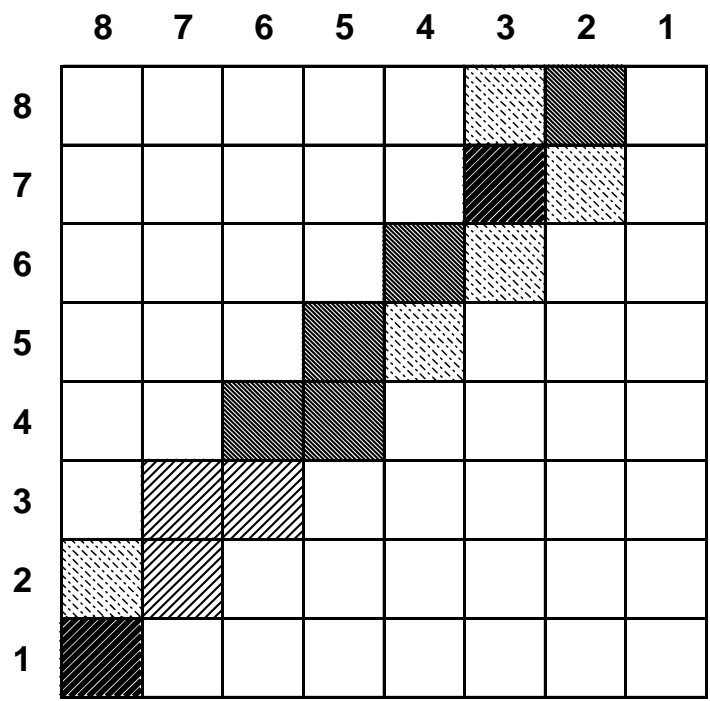

**Fig. 22**